\documentclass[sigconf]{acmart}
\usepackage[utf8]{inputenc}
\usepackage{hyperref}
\usepackage{listings}
\usepackage{color}
\usepackage{subcaption}
\usepackage{graphicx}
\usepackage{soul}
\usepackage{todonotes}
\usepackage{algorithm}
\usepackage{algorithmic}
\usepackage{xspace}
\usepackage{cleveref}
\usepackage{amsthm}
\usepackage{tikz}
\usepackage{pifont}
\usepackage{booktabs}
\usepackage{array}

\graphicspath{{figures/}}

\newcommand{\ie}{{\em i.e.,\/ }}
\newcommand{\eg}{{\em e.g.,\/ }}

\newcommand{\etal}{{\em et al.\/}}

\theoremstyle{definition}

\newcommand*\circled[1]{\tikz[baseline=(char.base)]{
            \node[shape=circle,draw,inner sep=0.1pt] (char) {\textbf{#1}};}}
\newcommand*\circledblk[1]{\tikz[baseline=(char.base)]{
            \node[shape=circle,draw,inner sep=0.1pt,fill=black,text=white] (char) {\textbf{#1}};}}
\newcommand*\circledblkr[1]{\tikz[baseline=(char.base)]{
            \node[shape=circle,draw,inner sep=0.1pt,fill=black,text=white,rotate=90] (char) {\textbf{#1}};}}

\newcommand\mustafa[1]{}
\newcommand\alberto[1]{}
\newcommand\yiannis[1]{}
\newcommand\michal[1]{}

\newcommand\sysname{Airtnt\xspace}

\newcommand\function{p}
\newcommand\teefunction{w}
\newcommand\hash{h}

\newcommand\outbuffer{\vec{o}}
\newcommand\encryptedoutbuffer{\{\vec{o}\}_{k}}

\newcommand\sstate{\mathsf{s}}
\newcommand\cycles{\mathsf{c}}
\newcommand\cyclesdone{\mathsf{d}}

\newcommand\encryptednewstate{\{\mathsf{s}'\}_{k}}
\newcommand\key{\mathsf{k}}

\begin{document}

\author{Mustafa Al-Bassam}
\affiliation{University College London}
\email{m.albassam@cs.ucl.ac.uk}
\author{Alberto Sonnino}
\affiliation{University College London}
\email{alberto.sonnino@ucl.ac.uk}
\author{Michał Król}
\affiliation{University College London}
\email{m.krol@ucl.ac.uk}
\author{Ioannis Psaras}
\affiliation{University College London} 
\email{i.psaras@ucl.ac.uk}

\begin{abstract}

We present \sysname, a novel scheme that enables users with CPUs that support Trusted Execution Environments (TEEs) and remote attestation to rent out computing time on secure enclaves to untrusted users. \sysname makes use of the attestation capabilities of TEEs and smart contracts on distributed ledgers to guarantee the fair exchange of the payment and the result of an execution. \sysname makes use of off-chain payment channels to allow requesters to pay executing nodes for intermediate ``snapshots" of the state of an execution. Effectively, this step-by-step ``compute-payment" cycle realises untrusted pay-as-you-go micropayments for computation. Neither the requester nor the executing node can walk away and incur monetary loss to the other party. This also allows requesters to continue executions on other executing nodes if the original executing node becomes unavailable or goes offline.

\end{abstract}

\title{\sysname: Fair Exchange Payment for Outsourced Secure Enclave Computations}

\maketitle

\section{Introduction}

The cloud computing model developed during the last two decades and realised by cloud computing providers was built on the premise of compute centralisation. That is, computing power is geographically and administratively concentrated in compute infrastructures of industrial scale, generally called datacenters.
The centralisation of computing comes with several drawbacks.
Firstly, in a world with an increasing number of devices that have low compute and storage resources, or produce enormous amounts of data \cite{autonomous-cars} at the edge of the network (\eg IoT devices, autonomous cars, or Chromebooks), computation inevitably needs to be outsourced to external computation nodes. This increased amount of data that needs to be transferred from the edge of the network towards the core (to remote servers in order to be processed) is expected to put significant strain on ISPs, inter-ISP business relationships and the Internet backbone.
Recent research has thus been focused on finding ways to bring the servers to the edge, so that most information can stay within the domain boundaries of the edge-ISP rather than having to traverse the Internet backbone \cite{8030322,7980109}.

Secondly, cloud computing services act as large central points of failure--for example, Amazon Web Services (AWS) controls up to 40\% of the cloud-server market, and when AWS's Virginia datacenter had an outage, a significant part of the web was offline~\cite{awsdown}. These central points of control also make it possible for authorities to enforce censorship of specific uses of the cloud, canceling any censorship-resistant property~\cite{awscensor}.

There is therefore a pressing need to develop alternative, decentralised computing infrastructures. To do so, the features of centralisation need to be re-engineered to fit into a decentralised and distributed computing domain. For instance, in the centralised model of cloud computing, the user and the server provider trust not to defraud each other (\ie the provider will provide the services that the user paid for, and the user agrees to provide the payment). The user also trusts the server provider to not tamper with the task that the user would like to execute, or its resulting data. The cloud provider achieves that by building reputation arounds its services, while disputes (between users and cloud providers) are resolved through the (physical) court system.

In a decentralised setting, instead, trust and reputation is more difficult to build as any node can join the system, provide services and get paid. That said, users do not know who to trust. Building a reputation system in this case, 
requires a trusted third party to enforce a ``one review per user" rule to prevent Sybils~\cite{douceur2002sybil}. An important research task is therefore, the following: \emph{``How can we design a decentralised computing platform where executing nodes can execute user's tasks, and receive payment for it (`fair exchange'), without: i) the user and the node trusting each other, ii) a third party to settle disputes and iii) possibility for either party to defraud each other (\eg by lack of payment from the user side or incorrect execution from the compute provider side''},

Trusted Execution Environments (TEEs)~\cite{costan2016intel,sgx1,sgx2}, have the potential to address the execution integrity issue, by enabling secure communication between the virtual instance of an application and an external entity, as well as tamper-proof execution (see~\cref{sec:sgx}). However by themselves they do not allow for the fair exchange of payment and result for two mutually distrusting parties.

Fair exchange schemes have traditionally required the participation of a trusted third party~\cite{pagnia1999impossibility}. However, trustless distributed ledgers (`blockchains') and smart contracts~\cite{buterin2013ethereum} (see Section~\ref{sec:smart_contract}) can take the role of the mediating third party instead, and facilitate the fair exchange of the payment and execution result. By augmenting TEEs so that results attested by the TEE can be fed to a smart contract, fair exchange can be facilitated.

One of the key challenges of this approach however, is to overcome the lack of practicality in sending large results to a smart contract, which all nodes have to verify. Storing large amounts of data in a blockchain is associated with high transaction fees~\cite{ew}. It would therefore be desirable to have to avoid to store the result at all in the blockchain.

We make the following contributions. 

\begin{itemize}
\item We design a fair exchange system, called \sysname, for TEE computation results and payments, according to which a malicious user is not able to defraud others and make them lose money, even if the malicious user is willing to lose money.
\item We design all the required protocols and protocol features to support \textit{execution integrity} (\ie requested computations should be executed correctly without anyone being able to tamper with the execution, and the requester should have a cryptographic proof of the correct execution).
\item We design a micropayment system using payment channels for requesters to pay executing nodes as they execute the computation. Requesters receive the changes to the state of the computation after every micropayment and executing nodes receive the micropayment in order to continue to the next state of the computation. This way, the computation can also continue on a different executing node should the original one go offline.
\item We implement and evaluate a prototype to demonstrate the practicality of our solution. We implement two separate use-cases, namely, Optical Character Recognition and Game of Life simulations, each of which demonstrates a separate feature of \sysname.
\end{itemize}

\section{Simplified System Overview}

\begin{figure}[t]
    \centering
    \includegraphics[width=.4\textwidth]{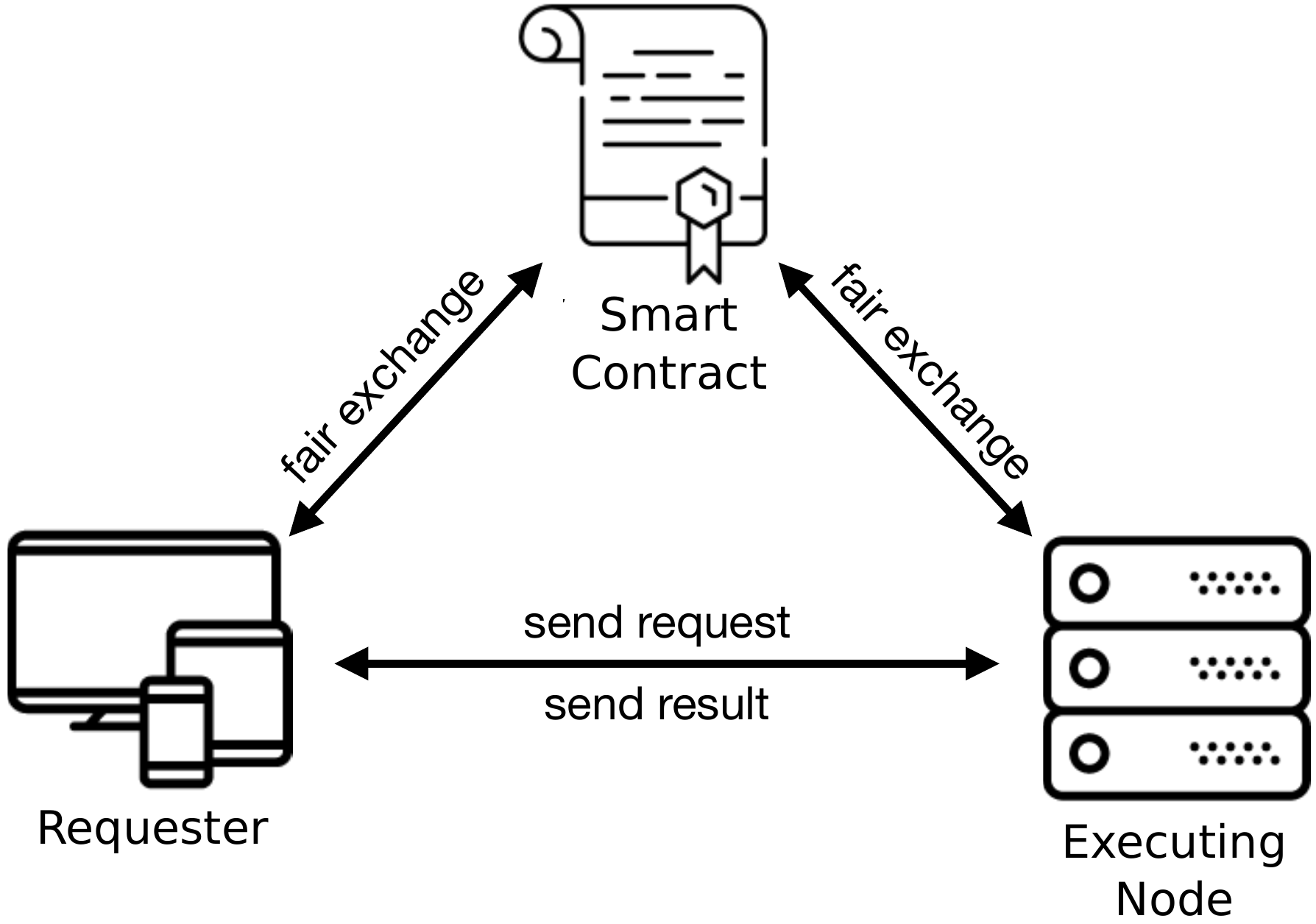}
    \caption{Simplified \sysname overview.}
    \label{fig:overview}
\end{figure}

In \sysname, we consider that executing nodes receive requests from requesters to execute predefined functions on some inputs. The mechanism for requesters and executing nodes to discover each other is out of the scope of this paper (\eg it could work as a web-based service that acts as a marketplace where requesters can bid for executing nodes to execute their requests).

Upon executing a request, an executing node responds to the requester with the result encrypted with a newly generated ephemeral secret key, and the cryptographic hash of the secret key, both generated and attested by the TEE during the execution of the request. TEE attestation ensures that the executing node has computed the correct function and is sending back the correct result. The executing node then submits the secret key itself to a smart contract initialised by the requester. The contract is programmed to send funds to the executing node upon submission of the preimage of the hash of the secret key sent to the requester. This allows the user to decrypt the result, thus achieving the fair exchange of the money and the execution result without trusting a third party (the third party is effectively the smart contract, which facilitates the fair exchange of the payment and execution result, and is untrusted).

Additionally, \sysname allows for long executions to be `paused' after certain checkpoints in the execution (which we call `execution cycles'), returning the intermediate state of the execution of the function as the result to the requester. The requester can then continue the execution by submitting a micropayment through a special type of \sysname payment channel, off the blockchain (see~\cref{sec:channels}).
Micropayments reduce the impact of a requesting node going offline or becoming dishonest after a long execution and wasting the requesting node's resources without payment. It can be thought of as a way to `livestream computation' with micropayments. The executing node only continues the execution after a micropayment has been received.

The payment channel is closed at the end of the execution, or if there is a dispute. If the executing node goes offline at any point during the execution, the requester can continue the execution with a different executing node using the intermediate state.
\section{Background}
We present background on smart contracts, payment channels and trusted execution environments.
\subsection{Smart Contracts on Blockchains}
\label{sec:smart_contract}

The concept of a blockchain was first proposed in Bitcoin~\cite{nakamoto2008bitcoin}, as a decentralised append-only ledger of financial transactions. The Bitcoin blockchain provides a global ordering on the transactions, in order to prevent funds being spent twice (the `double-spend' attack). As by now there is extensive literature on this topic, we only mention the properties of blockchains that \sysname relies on.

Bitcoin transactions have a simple internal scripting language that allow the transaction creator to define, as a script, the recipient of the transaction, such that in order for the recipient to spend the funds in the transaction, they must provide an input that causes the script to return $true$. The most common Bitcoin script defines a hash of the public key for the recipient, and returns true upon an input that provides a valid signature associated with that public key. The hash of the public key is referred to as an `address', as senders can use it to send funds to the owner of the key.

Blockchain platforms such as Ethereum~\cite{buterin2013ethereum} have extended Bitcoin's script language to allow users to execute more complex programs on the blockchain, called `smart contracts'. These are interfaces that users can send funds to, such that the management of those funds are defined by the code of the smart contract. Smart contracts in Ethereum can be written in high-level languages (\ie Solidity, a JavaScript-like language for Ethereum smart contracts), and are compiled down to Ethereum Virtual Machine (EVM) assembly code. This is then published on the blockchain to create instances of smart contracts, which have their own addresses. Like classes in object oriented programming languages, instances of smart contracts have methods that can be called.

Executing a transaction calling a method in an Ethereum smart contract has a `gas' cost associated with it; the more assembly opcodes and storage the transaction has, the higher the gas cost. The price of gas varies depending on the load on the network, and is paid for using Ether--Ethereum's built-in currency.

Blockchain platforms such as Bitcoin and Ethereum use proof-of-work as a consensus mechanism to agree on the ordering of blocks (which contain batches of transactions that update the state of the ledger), where nodes called miners create new blocks by repeatedly hashing the block until it is below a target value, which is adjusted by the network such that a block is generated every 10 minutes in the case of Bitcoin, or 30 seconds in the case of Ethereum. In the case of a fork, the chain with the most accumulated proof-of-work is considered the correct chain.
The security model for proof-of-work blockchains is that in order for a party to undo a transaction, they must create a fork of the chain with more accumulated proof-of-work that the chain that has the transaction in it. In order to do so, they need to be able to expend more mining power than the rest of the network (the `51\% attack'), thus making such attacks economically unpractical.


\subsection{Payment Channels}
\label{sec:channels}

The fact that all nodes must verify every transaction in public blockchains, such as Bitcoin or Ethereum, raises scalability issues and limits their usability in practice~\cite{bitcoinfees}. Especially for systems such as \sysname, where micropayments are needed, the total transaction fees would become prohibitive.

As a result, research in this area has shifted to \emph{off-chain payments}~\cite{offchain} using payment channels~\cite{paymentchannels}. This allows two parties to make payments to each other without recording all of their transactions on the blockchain.
This typically works by requiring both parties to create an initial transaction on the blockchain and open the channel by depositing a certain amount of coins. For as long as the channel is open, the two parties can make unlimited payments directly to each other without touching the blockchain, effectively updating each other's balance locally. The channel can then be closed by either party, settling the balance on the blockchain with a final transaction. This can be thought of as a similar process to opening a tab at a bar.

There are many technical proposals for designing payment channel systems on top of the Bitcoin blockchain~\cite{paymentchannels}. These proposals are designed so that neither of the two parties can steal funds from each other, or need to trust each other. In \sysname we focus on unidirectional payment channels using Ethereum, as they are simple to implement as a smart contract, without changes to the protocol.

Unidirectional payment channels only allow one party (the sender) to make payments to the other party (the recipient). In order to close a channel and settle a balance, both parties must sign a message with the owed amount to the recipient and submit it to the smart contract, which will then send the owed coins to the recipient.
The typical workflow of a unidirectional payment channel involves the sender sending a signed message to the recipient increasing the owed amount of coins. The channel is updated and the recipient performs the agreed service. Finally, the recipient closes the channel by signing the last update to the channel, and sending the two signatures to the smart contract, releasing the deposited funds.

To prevent the need for a user to have to open a payment channel and maintain a deposit with everyone that they would like to interact with, it is possible to create \emph{multi-hop payment channels}. This allows a payment to travel across multiple users, updating the balances of multiple payment channels. The Lightning network~\cite{lightning} is one example of this for Bitcoin, and the Raiden network~\cite{raiden} is an example for Ethereum.

\subsection{Trusted Execution Environments}
\label{sec:sgx}
Traditionally, one protects the integrity and confidentiality of applications by enforcing the isolation of applications. An operating system can isolate applications using hardware mechanisms like virtual address spaces and privileged instructions. Multiple operating systems running on the same physical host are isolated by the hypervisor using hardware virtualisation extensions provided by the CPU. However, \sysname assumes that functions are executed on untrusted nodes and that both the hypervisor and the OS can be compromised. At the same time, in \sysname, a running virtual machine requires a private key or a password to decrypt incoming data. The secret must remain confidential and protected against access from the hosting node. 

Trusted  Execution  Environment  (TEE)  is  an  environment  where code being executed inside the TEE is trusted in authenticity and integrity and both the code  and  other  assets  are  protected  from external access. Multiple TEEs are currently being developed for mobile devices \cite{ekberg2013trusted} and the most popular CPU architectures ~\cite{costan2016intel,winter2008trusted}

\paragraph{Intel Secure Guard Extension}

Intel SGX~\cite{costan2016intel,sgx1,sgx2} is an example of Trusted Execution Environment (TEE) that allows applications or part of applications' code to be executed in a secure container, called \textit{enclave}, protecting the integrity and confidentiality of code and data using hardware mechanisms directly in the CPU. SGX enclaves are protected from other applications, privileged system software, such as the operating system (OS), hypervisor, and BIOS, as well as attackers having physical access to the machine. SGX implements hardware encryption in the CPU so that only the authorised enclave can access its region of memory.
To enable an application to use enclaves, the developer must provide a signed shared library that will be executed inside an enclave. The library itself is not encrypted and can be inspected before being started, hence no secret should be stored inside the code. The enclave code cannot directly access OS functions (e.g., networking, I/O); it must invoke these functions through special entry-points that are under strict control of the enclave application.

\paragraph{Remote Attestation}

SGX also provides a remote attestation protocol. Using remote attestation an enclave can obtain a statement from the CPU attesting that it is running a particular enclave code with a given memory footprint. A requester can use this attestation to verify the identity and integrity of a target enclave running on a remote host, be convinced that the attestation has been generated by a legitimate SGX CPU, and securely transfer confidential data using an encrypted connection \cite{gueron2016memory}.
Intel SGX provides each enclave with a seal key that can be used to store data on permanent storage and access it again upon subsequent execution. This facilitates the development of applications that can restart an enclave without requiring a new remote attestation. The enclave instead loads its secrets from a configuration file encrypted with the enclave specific seal key and kept in stable storage. 

Currently, SGX supports only specially prepared Linux and Windows libraries (.so and .dll files), but multiple works extend it to legacy application and containers\footnote{Note that only the part of the application that is  processing sensitive data needs to run in the enclave---not the whole application.} \cite{baumann2015shielding,arnautov2016scone,shinde2017panoply}. In its current version, SGX is susceptible to cache-timing attacks \cite{Gotzfried:2017:CAI:3065913.3065915}, but an enhancement has already been proposed to eradicate this vulnerability \cite{shih2017t}. Moreover, SGX does not introduce significant overhead or increase of execution time \cite{zhao2016performance}.

\section{Threat Model and Goals}\label{threat_model}

The following actors participate in an \sysname transaction:

\begin{itemize}
\item \textbf{Requesters.} End-user (possibly constrained IoT) devices submit resource-heavy tasks to executing nodes. End-users need to pay the executing node for the CPU cycles that the latter has spent to complete the requested computation.
\item \textbf{Executing nodes.} High CPU capacity nodes using their resources to execute requested tasks/functions. Executing nodes are paid for carrying out computations.
\end{itemize}


We assume that both parties mutually distrust one another. Each party is potentially malicious, \ie they may attempt to steal funds, avoid making payments, and forge results, if it benefits them. At any given time, each party may drop, send, record, modify, and replay arbitrary messages in the protocol.

We assume that each executing node has a TEE-capable machine. Both the requester and the executing node trust their own environments, the TEE, and the function running in the enclave (\ie the function has been checked before by the user or the community). The enclave is thus trusted to correctly compute and attest the result and not leak information to the hosting execution platform. The rest of the system, such as the network between the parties and the other party’s software stacks (outside the TEE) and hardware are untrusted. During function execution, the executing node may therefore: (i) access or modify any data in its memory or stored on disk; (ii) view or modify its application code; and (iii) control any aspect of its OS and other privileged software.

\sysname makes use of smart contracts on a blockchain to facilitate payments \emph{without a trusted third party}. We assume that the underlying blockchain where the smart contract runs on is resistant to double-spending attacks, and has liveness - that is, transactions submitted to the blockchain will be eventually processed, within some defined period of time.

Given this threat model \sysname achieves the following goals:
\begin{itemize}

\item \textbf{Fair exchange.} A requester will only receive the result of an execution if the executing node gets paid for the execution, and vice versa: an executing node will only get paid for the execution, if the requester receives the result.

\item \textbf{Executing node counterparty risk resistance.} In the event of a requester going offline or diverging from the protocol, a requester cannot cause an executing node to perform a large amount of work without payment.

\item \textbf{Execution transferability.} In the event of an executing node going offline or diverging from the protocol, a requester should be able to resume a computation on a different executing node, without losing all of the work that the original executing node has already done on the computation.

\item \textbf{Execution integrity.} The result of computations returned to requesters must be correct and verifiable that the executing node executed the correct program.

\end{itemize}

\section{System Design}

We describe the \sysname execution model within a TEE, and how this model can be used for the fair exchange of i) payments and ii) computation results.

\subsection{Preliminaries and Notation}
We present the notion used in the rest of the paper, as well as the primitives on which \sysname relies.

\paragraph{Cryptographic Primitives.} \sysname assumes a cryptographically secure hash function $\hash(x)$. It also assumes a symmetric-key encryption algorithm and a random key generation function $generateKey()$. We denote a message $m$ encrypted with the symmetric key $k$ as $\{m\}_k$.

\paragraph{Smart Contract Primitives.} For parts of the scheme that rely on the public-key cryptography native to smart contracts, we denote a message $m$ encrypted with a private key $sk$ as $[m]_{sk}$. We assume a smart contract environment with support for the following primitives:
\begin{description}
\item [$\bullet$\xspace checkSig$(m, \sigma, addr)$:] returns $true$ if the signature $\sigma$ is a valid cryptographic signature of the message $m$ by address $addr$.
\item [$\bullet$\xspace send$(v, addr)$:] sends $v$ coins to the address $addr$.
\item [$\bullet$\xspace destroy$()$:] terminates the contract, so that no party can call any of the contract's functions.
\end{description}


\subsection{Execution Model}\label{sec:execution-model}
We present the \sysname execution model.

\paragraph{\sysname Functions.}
Requesters within \sysname can request executing nodes to execute \sysname functions, which are programs loaded in a secure enclave and remotely attested by the TEE. These programs are predefined; the mechanism by which a requester can distribute a program to executing nodes is irrespective of the \sysname protocol and out of the scope of this work.

A requester requests for an executing node to compute the result of an \sysname function $\function$. These functions take as input a state $\sstate$, and the number of execution cycles $\cycles$ to perform. They output the new state $\sstate'$, a set of messages representing the output of the program $\outbuffer$, and the number of execution cycles actually performed $\cyclesdone$. If the execution terminates before $\cycles$ cycles are performed, then $\cyclesdone < \cycles$.

\begin{equation}
\function(\sstate, \cycles) \rightarrow \sstate', \outbuffer, \cyclesdone
\label{eq:function}\end{equation}

In this model, state $\sstate_0$ of the function, provided in the beginning, also contains the initial input parameters to the function. However, the function may no longer need to work with the initial input parameters in the middle of the execution, thus we do not explicitly pass them to the function.

The demarcation of execution cycles is arbitrarily defined by the implementation of a function $p$---as it is the function that is responsible for calculating the number of cycles done $\cyclesdone$---and does not necessarily correspond to execution cycles on the underlying CPU\footnote{It is not always practical to count the number of instructions executed in a program during runtime, and CPU may not be the only resource requirements of a program---some programs may for example be more bandwidth- or memory-intensive.}. In that sense, each execution cycle acts as a checkpoint. Execution cycles in \sysname play a similar role to gas costs in Ethereum--they are a way to determine the cost of an execution. However an execution cycle on a non-terminal state always results in a new state, so unless $\sstate$ is a terminal state, $\sstate \neq \sstate'$ if $\cycles > 0$ in \Cref{eq:function}.

If execution of $\function$ is split up into multiple steps, it should give the same result and use the same number of execution cycles, such that given $\function(\sstate_0, a) \rightarrow \sstate_1, \outbuffer_1, \cyclesdone_1$ and $\function(\sstate_1, b) \rightarrow \sstate_2, \outbuffer_2, \cyclesdone_2$, then $\function(\sstate_0, a+b) \rightarrow \sstate_2, \outbuffer_1\cup\outbuffer_2, \cyclesdone_1+\cyclesdone_2$.

If after execution of $\function$, $\sstate'$ is not a terminal state resulting from $\function$ on $\sstate$, then the requester may feed $\sstate'$ into $\function$ again to continue the execution. Note that \sysname's novel design allows for the computation to be continued on a different executing node (\eg in case the original one goes offline).

\paragraph{Wrapper Function}
We also define a wrapper function $\teefunction$ for $\function$ that is executed by the TEE upon receiving a request from a requester. The wrapper function forms the basis of \sysname's fair exchange protocol. The function takes in as input an \sysname function $\function$, input state $\sstate$, and number of execution cycles to perform, $\cycles$. The function outputs the new state $\sstate'$ encrypted with an ephemeral symmetric key $\key$ ($\encryptednewstate$), the output buffer encrypted with $\key$ ($\encryptedoutbuffer$), the number of cycles performed $\cyclesdone$, and the cryptographic hash of $\key$ ($\key_{hashed}$). The value of these outputs are all attested by the TEE.

\begin{equation}
\teefunction(\function, \sstate, \cycles) \rightarrow \encryptednewstate, \encryptedoutbuffer, \cyclesdone, \key_{hashed}
\end{equation}

The wrapper function $\teefunction$ is the same for any function $\function$, thus a developer writing an application for \sysname is only concerned with defining $\function$ as their program.

\begin{algorithm}
\caption{Do procedure $\teefunction(\function, \sstate, \cycles)$}
\begin{algorithmic}
\STATE $\sstate', \outbuffer, \cyclesdone \leftarrow \function(\sstate, \cycles)$
\STATE $\key \leftarrow generateKey()$
\STATE $\key_{hashed} = \hash(\key)$
\RETURN $\encryptednewstate, \encryptedoutbuffer, \cyclesdone, \key_{hashed}$
\end{algorithmic}
\end{algorithm}

\paragraph{State Diffs}\label{sec:statediffs}

In many applications, the difference between a given state $\sstate$ and a later state $\sstate'$ may be small, \ie only a part of the state may change. It would therefore be wasteful, \eg in terms of bandwidth cost, for the executing node to have to resend parts of an intermediate state that the requester already has knowledge (an identical copy) of. Instead of the executing node sending the whole state to the requester, in \sysname, the executing node sends a \textit{diff} of the new state to the previous one containing only the changes. We assume a function $\textit{genDiff}$ that can be used by executing nodes to generate such diffs from two states, and a function $\textit{applyDiff}$ that can be used by requesters to decode the new state from a diff. The execution model is independent of the actual implementation of these functions.

\begin{equation}
\textit{genDiff}(\sstate, \sstate') \rightarrow sd
\end{equation}
\begin{equation}
\textit{applyDiff}(\sstate, sd) \rightarrow \sstate'
\end{equation}

The wrapper function can then be augmented to return the diff of the new state rather than the entire new state.

\begin{algorithm}
\caption{Do procedure $\teefunction(\function, \sstate, \cycles)$}
\begin{algorithmic}
\STATE $\sstate', \outbuffer, \cyclesdone \leftarrow \function(\sstate, \cycles)$
\STATE $\key \leftarrow generateKey()$
\STATE $\key_{hashed} = \hash(\key)$
\RETURN $\{genDiff(\sstate, \sstate')\}_k, \encryptedoutbuffer, \cyclesdone, \key_{hashed}$
\end{algorithmic}
\end{algorithm}

\subsection{Payments Protocol}\label{sec:payments-protocol}

We describe \sysname's payments protocol based on the execution model described above. The protocol builds on a smart contract-based unidirectional payment channel between the requester and the executing node.

Firstly, we define a smart contract for creating and managing an \sysname-specific unidirectional payment channel that enables the requester to send micropayments to the executing node after each intermediate state transition. The contract is held at $id_{contract}$ and is initiated by the function $initChannel(addr_r, addr_e, timeout, deposit)$ containing the following variables:

\begin{itemize}

\item $addr_r$: the address of the of requester (corresponding to a public key);

\item $addr_e$: the address of the executing node (corresponding to a public key);

\item $timeout$: the date that the channel will expire;

\item $deposit$: the number of coins that the requester has deposited into the contract.

\end{itemize}

Additionally, corresponding to addresses $addr_r$ and $addr_e$ are private keys $sk_r$ and $sk_e$.

After the contract is initialised, the balance (\ie number of coins owed to the executing node) of the channel can be updated by the requester. To do so, the requester is sending: \emph{i)} a signature to the executing node for the hash of the message $id_{contract} || v || \key_{hashed}$, \emph{ii)} the concatenation of the ID of the contract $id_{contract}$, \emph{iii)} the total amount of coins owed to the requester $v$, and \emph{iv)} the hash of the secret key $\key_{hashed}$ provided by the executing node after the last execution round (see \Cref{sec:execution-model}). Note that this procedure takes place off the smart contract platform, as the requester sends the signature directly to the executing node. We denote the new channel state signed by a private key $sk_x$ as follows:

\begin{equation}
[\hash(id_{contract} || v || \key_{hashed})]_{sk_x}
\end{equation}

The executing node can call the smart contract method $closeChannel$ by presenting a valid signature for the new state of the channel from the requester, as well as its own signature for the state of the channel (and the new balance of the channel). Crucially, the executing node must provide to the smart contract the pre-image $\key$ of $\key_{hashed}$ in order to close the channel and receive the funds, so that the requester can decrypt the computation result, as will be discussed in~\Cref{sec:protocol}.

\begin{algorithm}
\caption{Do procedure $closeChannel(sig_r, sig_e, v, \key_{hashed}, k)$}
\begin{algorithmic}
\STATE $statehash \leftarrow \hash(id_{contract} || v || \hash(\key))$
\STATE assert $checkSig(statehash, sig_r, addr_r)$
\STATE assert $checkSig(statehash, sig_e, addr_e)$
\STATE assert $\key_{hashed} = \hash(k)$
\STATE assert $v <= deposit$
\STATE $send(v, addr_e)$
\STATE $send(deposit - v, addr_r)$
\STATE $destroy()$
\end{algorithmic}
\end{algorithm}

The payment channel also employs a \emph{timeout}, such that if the timeout is reached and the channel has not been closed, the channel can be destroyed and deposits are returned back to the senders. This is to prevent the case where the recipient of the channel (executing node in our case) is malicious and extorts the sender of the channel (requester in our case) to transfer extra money (without doing any work) or otherwise, risking losing their deposit. The malicious receiver/executing node can achieve that by just refusing to close the channel. If this happens, the requesting node can simply wait until the timeout to recover the deposited funds.

We thus define a smart contract procedure $channelTimeout$ that can be called after the timeout is reached.

\begin{algorithm}
\caption{Do procedure $channelTimeout()$}
\begin{algorithmic}
\IF{$timeout > now$}
\STATE $send(deposit, addr_r)$
\STATE $destroy()$
\ENDIF
\end{algorithmic}
\end{algorithm}

\subsection{\sysname protocol}
\label{sec:protocol}

\begin{figure*}
\centering
\includegraphics[width=0.55\linewidth,keepaspectratio]{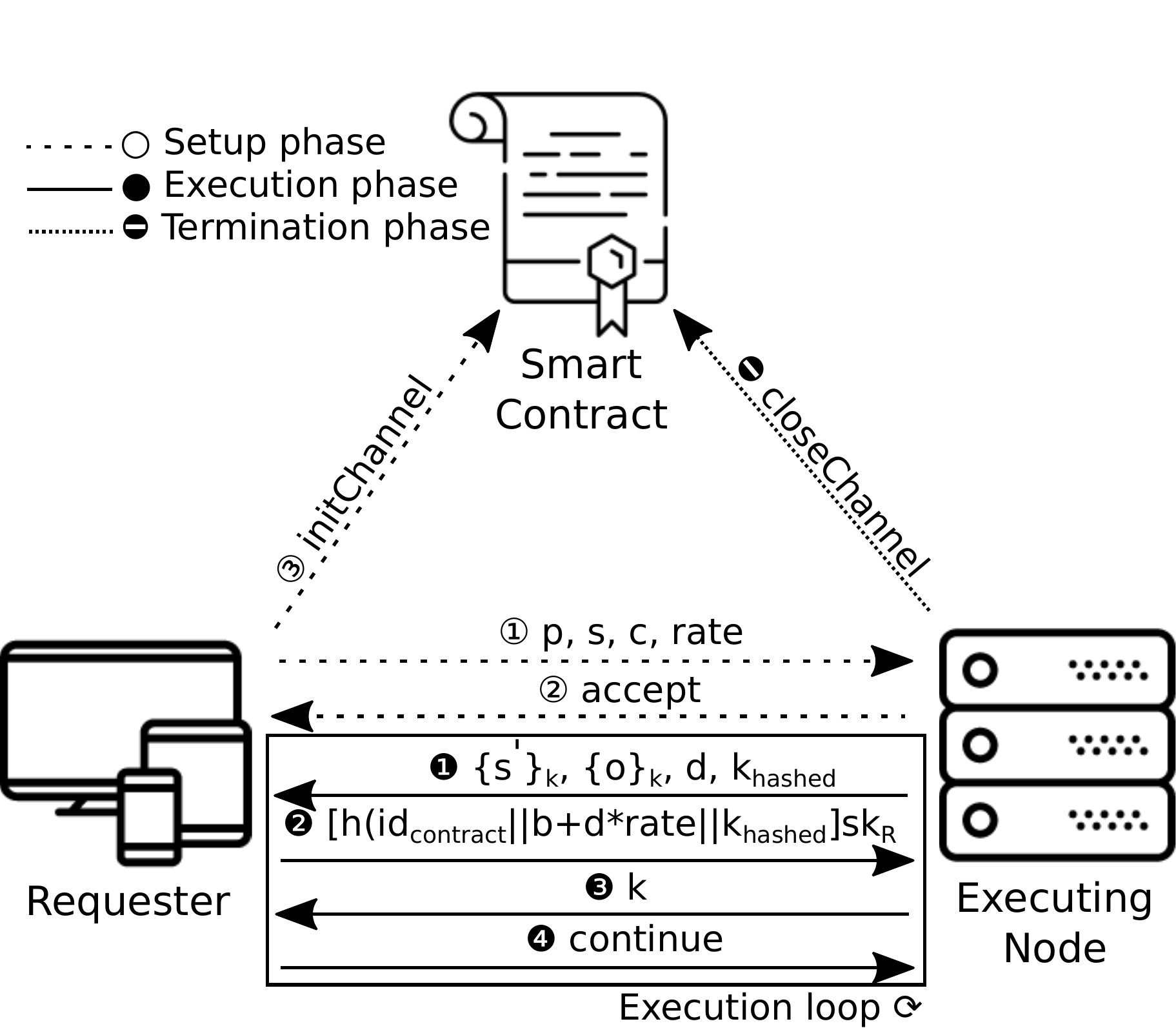}
\caption{Overview of interactions between a requester, an executing node and a smart contract in \sysname.}
\label{fig:design}
\end{figure*}

We describe the overall \sysname protocol unifying the execution model described in \Cref{sec:execution-model} and the payments protocol described in \Cref{sec:payments-protocol}. \Cref{fig:design} provides an an overview of the protocol.

\begin{description}
\item[Setup phase.] This phase is performed once at the beginning of every relationship between a requester and an executing node for a specific \sysname function.

\begin{itemize}

\item[\circled{1}] A requester $R$ sends $\function, \sstate, \cycles$ (the function to be executed, the initial state, and the number of cycles to perform) to an executing node $E$. $R$ also specifies some payment rate per cycle, $rate$, that defines the number of coins that $R$ is willing to pay $E$ for each computational cycle performed.

\item[\circled{2}] $E$ responds to $R$ by either accepting or rejecting the request.

\item[\circled{3}] If $E$ agrees to fulfill the request, then $R$ initializes a payment channel at $id_{contract}$, with the parameters $initChannel(\allowbreak addr_R, addr_E, timeout, deposit)$ where $addr_R$ is the address of the requester, $addr_E$ is the address of the executing node, $timeout$ is some conservative timeout that is longer than the time expected to fulfil the request, and $deposit$ is a value greater or equal to $rate * \cycles$.
If the payment channel is not initialised with the appropriate parameters, $E$ does not participate, and $R$ can proceed to the termination phase.
Both parties also locally maintain a variable $b$ which represents the balance of the channel so far, and is iniitalised with $b = 0$.

\end{itemize}

\item[Execution phase.] This is the phase where the main execution loop takes place.

\begin{enumerate}

\item[\circledblk{1}] $E$ executes $\teefunction(\function, \sstate, \cycles)$ in the TEE, returning $\encryptednewstate, \encryptedoutbuffer, \cyclesdone,\allowbreak \key_{hashed}$, which are all attested by the TEE, and sends $R$ these outputs.

\item[\circledblk{2}] $R$ checks that these outputs are attested by the TEE. If they are, then $R$ signs a new update to the payment channel, [$\hash(id_{contract} || b + \cyclesdone * rate || \key_{hashed})]_{sk_R}$, and sends it to $E$. $R$ then locally updates $b$ to $b + \cyclesdone * rate$.

\item[\circledblk{3}] Upon $E$ verifying that the signature received from $R$ is valid for the message $\hash(id_{contract} || b + \cyclesdone * rate || \key_{hashed})$, $E$ sends $R$ the secret key $\key$. Note that $E$ does not need to receive the message itself, as it already has the information to compute it. $E$ then locally updates $b$ to $b + \cyclesdone * rate$.
$R$ then decrypts $\encryptednewstate$ and $\encryptedoutbuffer$ to get $\sstate'$ and $\outbuffer$.

\item[\circledblk{4}] $R$ can now either request $E$ to continue the execution of $\function$ on the new state $\sstate'$, or terminate the protocol and find a different executing node to continue the execution on $\sstate'$, if $\sstate'$ is not the terminal state.

\end{enumerate}

\item[Termination phase.] There are two ways to terminate the protocol.

\begin{itemize}

\item[\circledblkr{\textsf{l}}] $E$ can close the payment channel by calling $closeChannel(\allowbreak sig_R,sig_E, v, \key_{hashed}, k)$, thus transferring the owed amount $v$ to $E$ and terminating the contract.
\item[\circledblkr{\textsf{l}}] If $E$ becomes unresponsive or goes offline, then $R$ can wait until the timeout, and call $channelTimeout()$ to get its deposit back.

\end{itemize}

\end{description}

\subsection{Evaluation}

We evaluate how the design of \sysname meets the threat model and design goals set out in~\Cref{threat_model}.

\paragraph{Fair Exchange.} \sysname aims to facilitate the fair exchange of payments and execution results. We break up this goal into two security properties, that we argue for: i) $E$ will only get paid iff $R$ receives the result of the execution and ii) $R$ will only receive the result of an execution iff $E$ can get paid for it. For the latter goal we say "can" rather than "will", because it is $E$'s responsibility to close the payment channel and collect the payment, and so if $E$ does not do this then they can voluntarily forfeit the payment, technically speaking.

\begin{theorem}
Assuming $R$ follows the protocol, an executing node $E$ will only get paid for executing $\teefunction(\function, \sstate, \cycles) \rightarrow \encryptednewstate, \encryptedoutbuffer, \cyclesdone, \key_{hashed}$ if and only if the requester $R$ receives $\sstate', \outbuffer$.
\begin{proof} (Informal.)
The only $send$ call in the smart contract that sends any coins to $addr_E$ is in the $closeChannel$ function. For $b$ coins to be sent to $addr_E$ via $closeChannel$, both the $E$ and $R$ must produce a signature of a message (the state hash) that authorises the transfer of $b$ to $addr_E$. However, assuming $R$ follows the protocol, then $R$ will only produce such a signature if and only if it receives $\key$ from $E$ to successfully decrypt $\encryptedoutbuffer, \cyclesdone$  with a correct attestation from the TEE. Furthermore, under the threat model described in \Cref{threat_model}, the TEE is trusted to correctly execute $w$ and produce the encryptions of $\encryptedoutbuffer, \cyclesdone$; only the TEE is capable of producing such attestation.
\end{proof}
\end{theorem}

\begin{theorem}
Assuming $E$ follows the protocol and can submit a $closeChannel$ transaction that is executed before the channel timeout, $R$ can only receive $\sstate', \outbuffer$ for an execution $\teefunction(\function, \sstate, \cycles) \rightarrow \encryptednewstate, \encryptedoutbuffer, \cyclesdone,\allowbreak \key_{hashed}$ if and only if $E$ can get paid for the execution.
\begin{proof}
(Informal.) Assuming $E$ follows the protocol, $E$ will only send $\key$, which is necessary to derive $\sstate', \outbuffer$ from $\encryptednewstate, \encryptedoutbuffer$, to $R$ if and only if $E$ receives from $R$ a signed update to the channel for the new balance $b$ that pays for the execution.
This signed update to the channel state enables $E$ to send a $closeChannel$ transaction to the blockchain to get paid $b$ coins, if it is called before the channel timeout. Under the threat model described in \Cref{threat_model}, the underlying blockchain network is assumed to be resistant to double spend attacks and will process the underlying transaction before the timeout (has liveness).
\end{proof}
\end{theorem}

\paragraph{Executing Node Counterparty Risk Resistance.}

We argue that $R$ can only cause $E$ to execute a single execution round of $\cycles$ cycles, without payment. Note that even if this happens, the fair exchange property of the protocol is not violated: $R$ must make a payment to receive the result.

\begin{theorem}
Assuming $E$ follows the protocol, $R$ cannot require $E$ to perform more than one execution $\teefunction(\function, \sstate, \cycles) \rightarrow \encryptednewstate, \encryptedoutbuffer, \cyclesdone,\allowbreak \key_{hashed}$ without updating the payment channel balance.
\begin{proof}
(Informal.) If $E$ executes $\teefunction$ and sends to $R$ $\encryptednewstate, \encryptedoutbuffer, \cyclesdone,\allowbreak \key_{hashed}$, but does not receive a signed channel update in response, $E$ will not execute $w$ on behalf of any more requests from $R$ until it receives a valid payment channel update, which may be never if the payment channel expires.
\end{proof}
\end{theorem}

\paragraph{Execution Transferability.}

This goal is met in step \circledblk{4} of the execution phase of the \sysname protocol, building on the execution mode described in~\Cref{sec:execution-model}. Upon decrypting the new intermediate state $\sstate'$ of the execution, $R$ can either request the same execution node to continue the execution on $\sstate'$, or find a different node to request the execution of the function on $\sstate'$.

\paragraph{Execution Integrity.}
The integrity of execution results rely on the security properties of the underlying TEE and on the cryptography underlying the remote attestation protocol. The TEE is trusted to correctly execute programs and produce attestations for their results. If, for example, the TEE has security vulnerabilities, the integrity of the results may be violated. Additionally, in the case of Intel SGX, Intel is a trusted third party, as they have the ability to use their cryptographic keys to generate fraudulent remote attestations.

Recall that in step \circledblk{2} of the execution phase of the \sysname protocol, $R$ must check that the outputs sent by $E$ are attested by the TEE, thus if the security properties of the TEE hold, then execution integrity also holds.
\section{Implementation and Performance} \label{implementation}

Recall in~\Cref{sec:statediffs} that executing nodes can return diffs of the new state, thus programs that manipulate state have different network overheads than programs that only consume state. We prototyped \sysname for those two types of applications which have different properties in terms of the way they handle state:

\begin{itemize}

\item \textbf{State-based programs (\ie simulations).} In simulations and other types of programs that rely heavily on the need to remember and manipulate intermediate state, \sysname executing nodes need to return the changes made to the state to requesters.

\item \textbf{Pure programs.} These programs do not need to modify state, but only process it and return output based on the state. After processing parts of the state, the program may discard the state if it does not need to revisit it. Intermediate states may be thus a subset of the input state, with only state removed and not added. These diffs would be negligible in size compared to state-based programs, as they only contain instructions to remove state at specific locations, and do not need to transfer new state data. In some cases, such as frame-by-frame image processing, these programs can be parallelized across multiple \sysname executing nodes.
\end{itemize}

We implement and evaluate a Game of Life simulation~\cite{conway1970game} to illustrate an example of a state-based program, and a simple Optical Character Recognition (OCR)~\cite{ocr} to illustrate an example of a pure program. We release both of these applications as open-source projects on GitHub\footnote{Removed for submission}. We instantiated the requester on Amazon AWS~\cite{aws} on a \textit{t2.xlarge} instance, and the executor on a Dell XPS laptop with Intel Core i7-8550U running Intel SGX~\cite{costan2016intel} as TEE.

\subsection{State-based programs} \label{sec:State-based programs}
To illustrate state-based programs we implement and evaluate a Game of Life simulation~\cite{conway1970game}. The Game of Life is a simple example of simulation acting on an arbitrarily large grid of cells, where each cell is either filled in or not. The game starts with some pattern of filled in cells, and evolves to obtain the next state by applying simple transition rules concurrently to each cell of the gird. We implement this example to show how \sysname can be used to outsource state-based simulations. 

\begin{figure}[t]
    \centering
    \includegraphics[width=.48\textwidth]{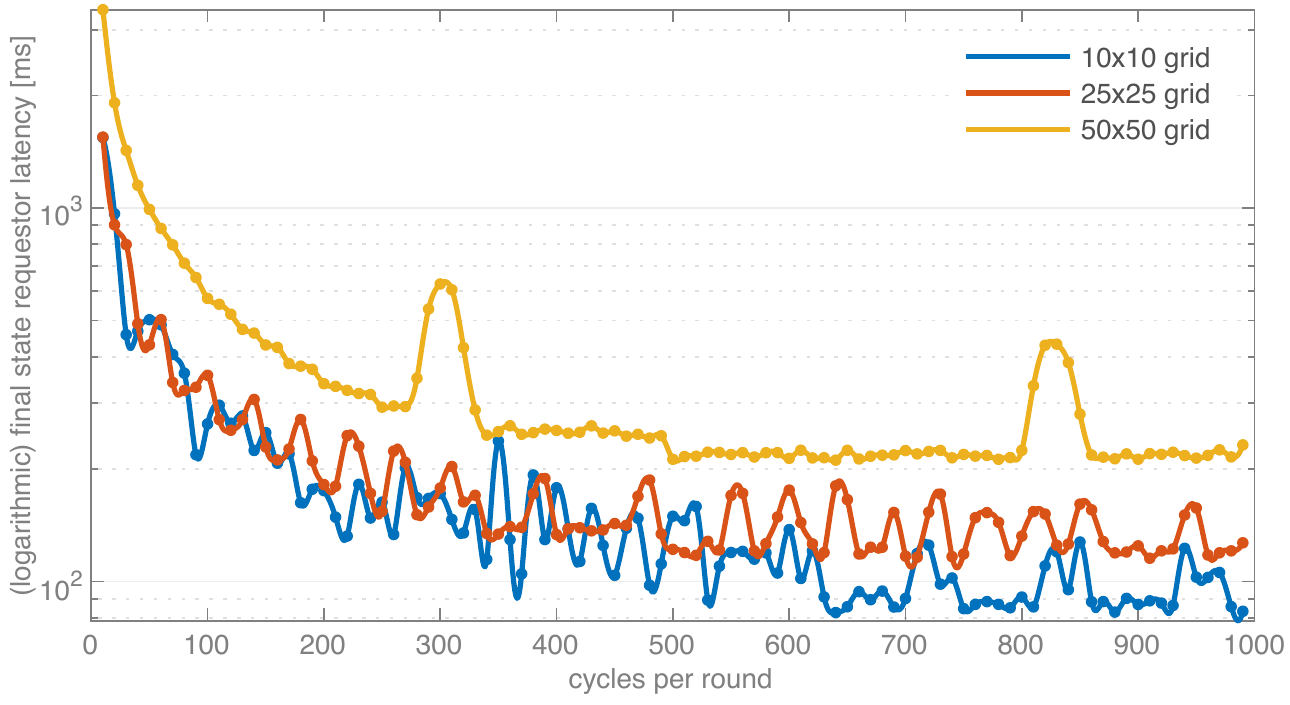}
    \caption{Game of Life -- variation of the client perceived latency of receiving the final state with the number of cycles per round for various grid sizes. The total number of cycles is fixed to 1000.}
    \label{fig:life_total_log}
\end{figure}

\begin{figure}[t]
    \centering
    \includegraphics[width=.48\textwidth]{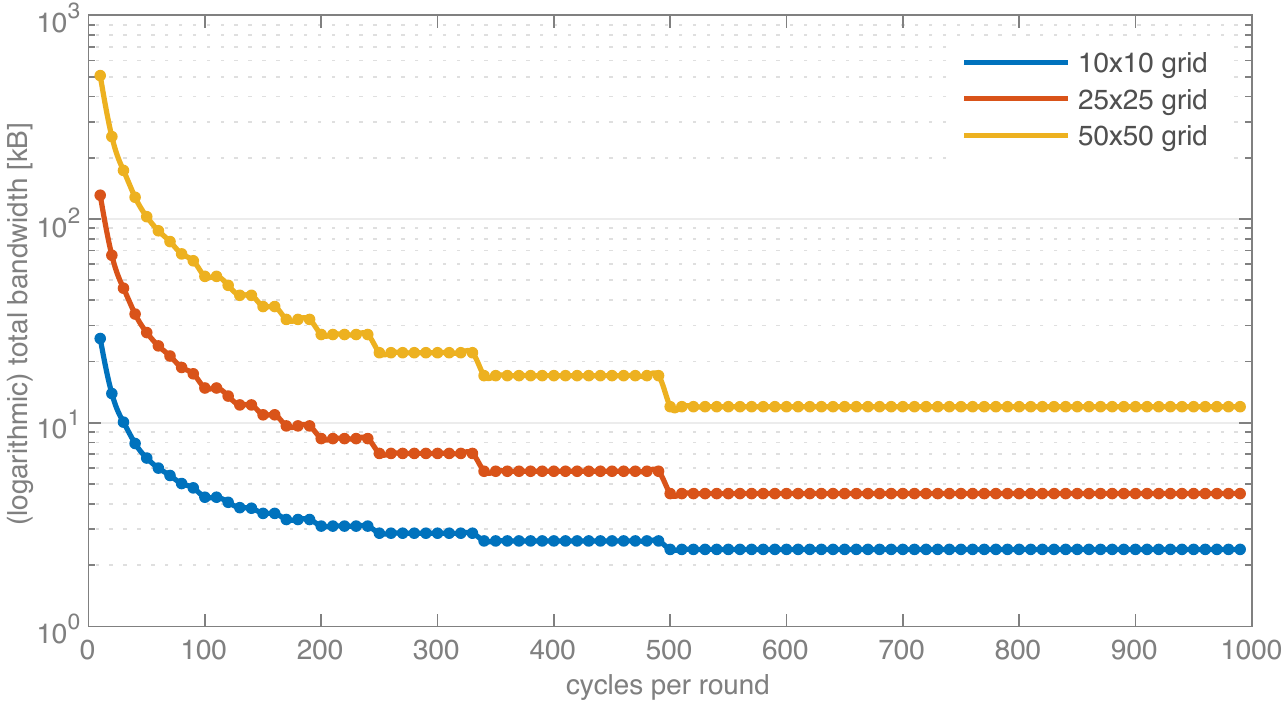}
    \caption{Game of Life -- variation of the bandwidth cost with the number of cycles per round for various grid sizes. The total number of cycles is fixed to 1000.}
    \label{fig:life_bandwidth_log}
\end{figure}

\Cref{fig:life_total_log} shows how the latency perceived by the requester (the time to receive the final state after first submitting the request) varies with the number of cycles per round (in semi-log scale); this graph captures the total execution time. We fix the total number of simulation steps to 1000 cycles ($c=1000$), and run the simulation for various grid sizes; $10 \times 10$, $25 \times 25$ , and $50 \times 50$ ($\sim 2.5 kB$ of data). As expected, the latency is much higher when requesting a few cycles per round---the requester has to transmit intermediate states back and forth every few cycles, which increases latency. When \sysname operates by steps of 10 cycles (leftmost point of the graph), the requester has to send and receive 100 intermediate states (the total number of cycles divided by the number of cycles per round) before completing the full simulation of 1000 cycles. However, when operating with number of cycles per round $\in[500,990]$, only two intermediate states are exchanged, and the latency is about 10 times lower. Moreover, \Cref{fig:life_total_log} shows that the latency first drops exponentially\footnote{Recall that the graph in \Cref{fig:life_total_log} is in semi-log scale.}  and then flattens out when increasing the number of cycles per round; increasing it from 10 to 200 reduces the latency by almost 10 times. This allows to achieve latencies of the order of $\sim 100 ms$, which is suitable for real-world scenarios. \Cref{fig:life_bandwidth_log} shows the variation of bandwidth cost with the number of cycles (in semi-log scale). As expected, the bandwidth cost decreases with the number of cycles; the number of intermediate states that the requester needs to send varies as the total number of rounds. The bandwidth cost is higher when the number of cycles per round is 10 as the requester sends 100 intermediate states, and then flattens out when the number of cycles per round $\in[500,990]$ as only 2 intermediate states are transmitted.  

\begin{figure}[t]
    \centering
    \includegraphics[width=.48\textwidth]{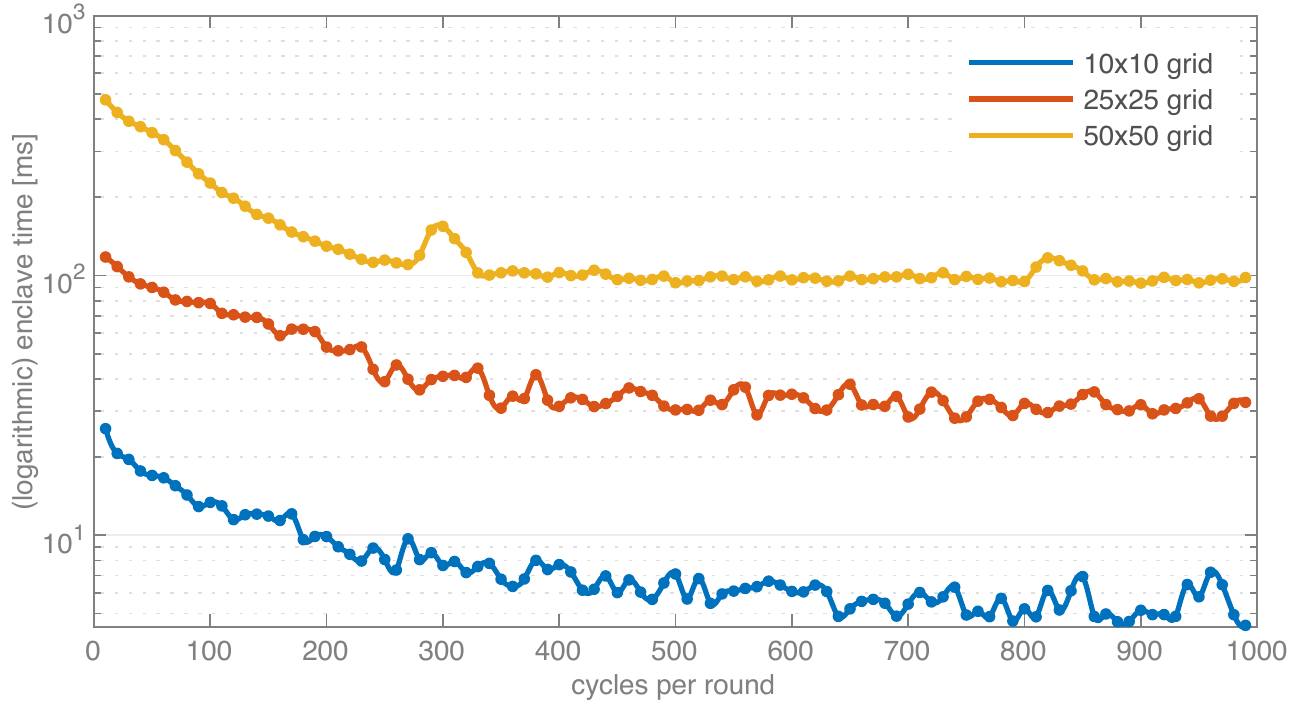}
    \caption{Game of Life -- variation of the enclave execution time with the number of cycles per round for various grid sizes. The total number of cycles is fixed to 1000.}
    \label{fig:life_enclave_log}
\end{figure}

\begin{figure}[t]
    \centering
    \includegraphics[width=.48\textwidth]{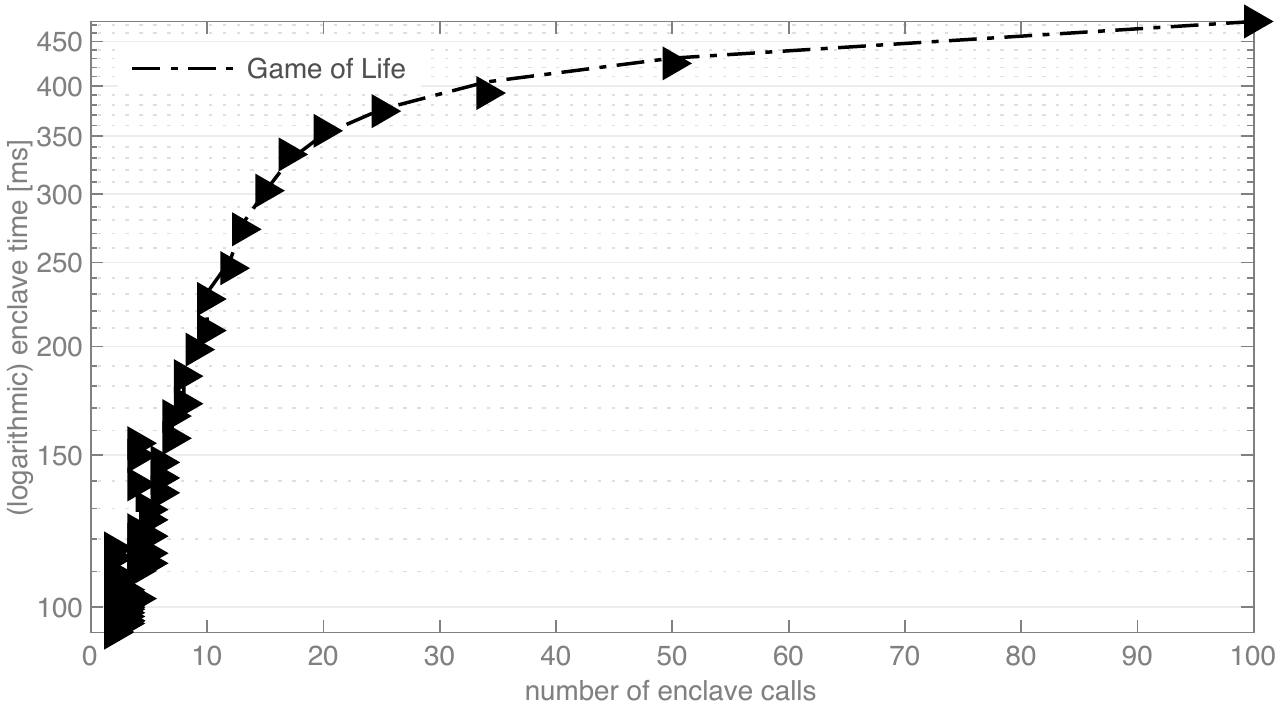}
    \caption{Game of Life -- variation of the enclave execution time with the number of enclave calls, for a $50 \times 50$ grid. The total number of cycles is fixed to 1000, and the number of enclave calls varies from 2 (for number of cycles per round $\in[500, 990]$) to 100 (for 10 cycles per round).}
    \label{fig:life_enclave_calls_log}
\end{figure}

\Cref{fig:life_enclave_log} shows the enclave execution time per cycle (in semi-log scale). The graph suggests that calling the SGX enclave multiple times introduces overhead. Therefore, working with low values for the number of cycles per round results in calling the enclave multiple times, which is expensive. To illustrate this phenomena, \Cref{fig:life_enclave_calls_log} shows the variation of the enclave execution time with the number of enclave calls. This graph shows that executing 1000 cycles with 2 calls (the number of cycles per round $\in[500,990]$) is about 5 times faster than executing them with 100 enclave calls (10 cycles per round). \Cref{fig:life_total_log} and \Cref{fig:life_enclave_log} also suggest that when the number of cycles per round is greater than 200 the communication time and the enclave execution time are of the same order of magnitude.

Our evaluation of Game of Life shows a clear trade-off between efficiency and low-risk. Executing \sysname a few cycles at the time (with low values for the number of cycles per round) helps with mitigating the risk of a requester dropping out without paying for the execution, but at the same time the cost of both bandwidth and execution time increase. However, our evaluations help to find the right balance; as noted above, increasing the number of cycles per round from 10 to 200 improves latency by about 10 times, while increasing it over 300 only adds a small benefit. Therefore, choosing 200 cycles per round seems to be a good choice for our implementation of Game of Life.

\subsection{Pure programs}

We implement and evaluate a simple Optical Character Recognition (OCR) algorithm~\cite{ocr} to illustrate the execution of pure programs. Our simple algorithm is initialized in the enclave with a model of each letter of the alphabet; it can then be fed with input images to detect the list of embedded letters. We implemented this example to show how \sysname can be used to outsource execution of pure programs.

\begin{figure}[t]
    \centering
    \includegraphics[width=.48\textwidth]{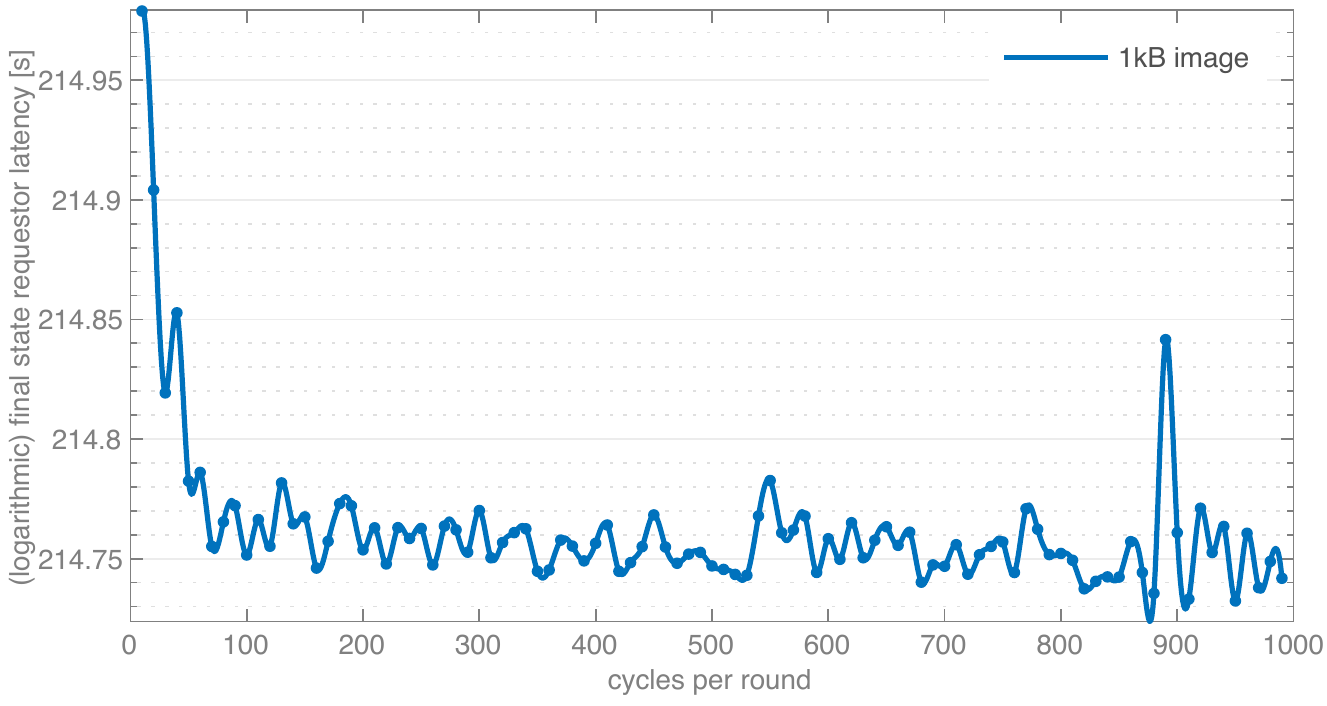}
    \caption{OCR -- variation of the client perceived latency on the number of cycles per round for images of $\sim 1kB$. The total number of cycles is fixed to 1000.}
    \label{fig:ocr_total_log}
\end{figure}

\begin{figure}[t]
    \centering
    \includegraphics[width=.48\textwidth]{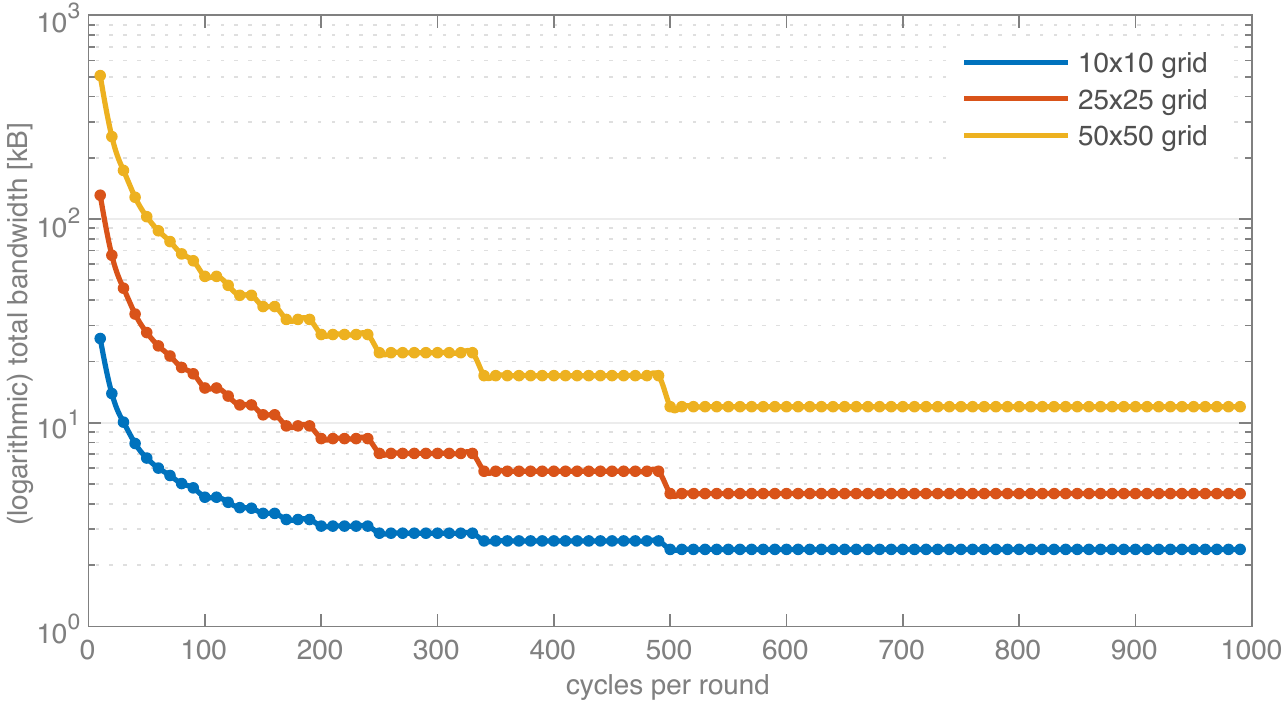}
    \caption{OCR -- variation of the bandwidth cost with the number of cycles per round for images of size $\sim 1kB$. The total number of cycles is fixed to 1000.}
    \label{fig:ocr_bandwidth_log}
\end{figure}

\Cref{fig:ocr_total_log} shows how the latency perceived by the requester varies with the number of cycles (in semi-log scale). We fix the total number of simulation steps to 1000 cycles in total as in \Cref{sec:State-based programs}, and run the simulation for images of size $\sim 1kB$. The latency decreases as for the Game of Life simulation but is much higher on overall because the execution of each cycle requires a transmission of a new input image. Therefore, the requester sends about 1.7MB of data to the executor for the processing of 1000 images (1000 cycles) of 1kB each; and the bandwidth cost stays constant regardless the number of cycles. 

\begin{figure}[t]
    \centering
    \includegraphics[width=.48\textwidth]{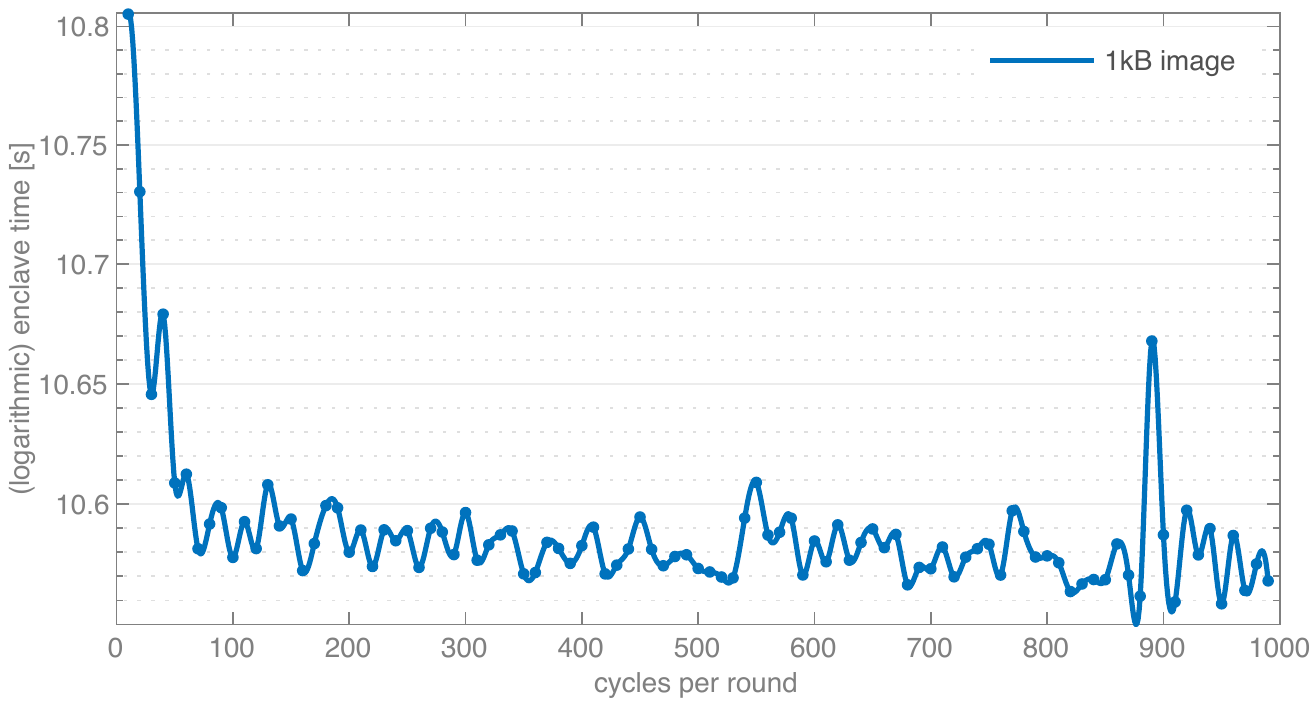}
    \caption{OCR -- variation of the enclave execution time with the number of cycles for images of size $\sim 1kB$. The total number of cycles is fixed to 1000.}
    \label{fig:ocr_enclave_log}
\end{figure}

\begin{figure}[t]
    \centering
    \includegraphics[width=.48\textwidth]{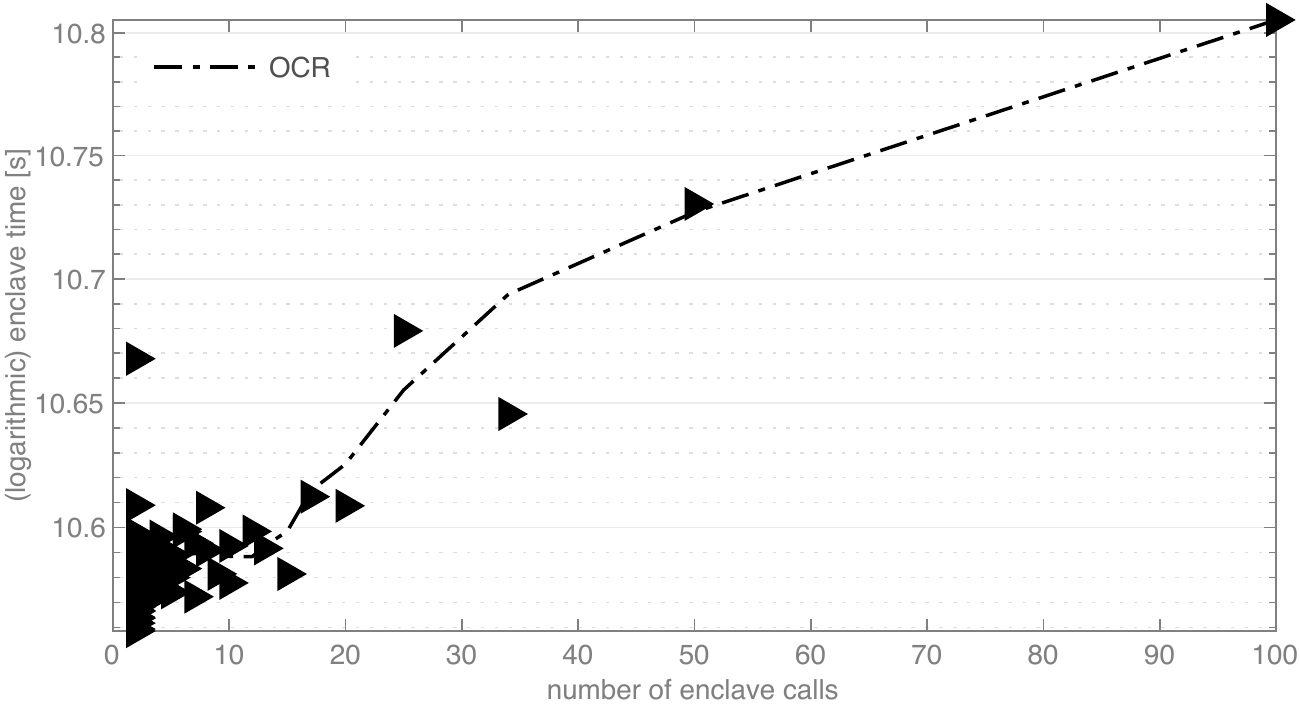}
    \caption{OCR -- variation of the enclave execution time with the number of enclave calls, for images of size $\sim 1kB$. The total number of cycles is fixed to 1000, and the number of enclave calls varies from 2 (for number of cycles per round $\in[500, 990]$) to 100 (for 10 cycles per round).}
    \label{fig:ocr_enclave_calls_log}
\end{figure}

\Cref{fig:ocr_enclave_log} and \Cref{fig:ocr_enclave_calls_log} respectively show the enclave execution time per cycles and per number of enclave calls (in semi-log scale). OCR is a much more CPU-intensive application than Game of Life---running the algorithm over 1000 images requires more than 10 seconds in the enclave. The enclave execution times decreases by 200 ms from 10 cycle per round to 100 cycles per round, and then stays roughly the same. As in \Cref{sec:State-based programs}, these graphs emphasize the overhead of making multiple calls to the SGX enclave. Moreover, \Cref{fig:ocr_total_log} and \Cref{fig:ocr_enclave_log} also suggest that the requester latency is dominated by the time to send the input images (and not by the enclave execution time).

Our experiments illustrate the differences between using \sysname to outsource the execution of state-based programs (e.g., Game of life) and pure programs (e.g., OCR). State-based programs can run multiple cycles from the same input as they operate on intermediate steps; this allows to save on bandwidth cost as it flattens out when increasing the number of cycles per round. Pure programs require a new input each cycle and cause the bandwidth cost to remain constant; the only benefit of increasing the number of cycles per round in this case is to reduce the number of enclave calls and thus save on execution time. However, pure programs can be parallelized. A requester can send a subset of inputs to many executors, and therefore divide latency by the number of executors involved in the computation.

\subsection{Limitations of Intel SGX}
In terms of performance, Intel SGX introduces memory limitations and performance overhead. Intel SGX limits the enclave memory to 128 MB; this limitation comes from the BIOS, and sets and upper bound to the input size that the executors can process simultaneously. It is however technically possible to extend that limit by editing the paging support~\cite{sgx2}. The performance overhead results from accessing the encrypted memory in an enclave and from the additional effort associated with entering and exiting an enclave. As shown in \Cref{implementation}, minimizing the number of accesses to the enclave significantly increases performance.

On the security side, relying on Intel SGX requires trust into Intel, as debated by many works recently~\cite{sgx_bad1, jackson2017trust, sgx_bad2, sgx_bad3}.

\subsection{Payment Channel Smart Contract}

We implemented an Ethereum smart contract, written in Solidity, for \sysname unidirectional payment channels.

Due to the Ethereum Virtual Machine having a maximum stack depth of 15, $closeChannel$ had to be implemented in such a way that it must be called twice: once for the requester's signature, and once for the executing node's signature. This is because the maximum stack depth limits the number of inputs a function may have. As the maximum size of a variable in the Ethereum Virtual Machine is 32 bytes, the amount of data that a function can take as input is too limited to accept two ECDSA signatures.

We present the gas costs incurred by different functions of the smart contract in~\Cref{table:contractcosts}. We note that the price of gas, and the market price of Ether itself, varies wildly from time to time due to volatility, so the USD cost is only accurate as of April 2018. The largest cost is creating the contract at \$0.46, as this involves uploading and storing the contract's code in the Ethereum blockchain. However, if the smart contract is uploaded as a library contract, this cost can be significantly reduced as the payment channel code only needs to be uploaded once.

The cost for initialising and closing a channel is \$0.10 and \$0.15 respectively, thus a complete \sysname transaction between a requester and executing node would cost \$0.25.
This cost could be reduced in the future by using multi-hop payment channels (see the Lightning network as an example~\cite{poon2015bitcoin}), so that a requesters do not need to open a payment channel with every executing node, as long as there is a path in the network between the requester and the executing node.


\newcolumntype{L}{l<{\hspace{1cm}}}
\newcolumntype{C}{c<{\hspace{.3cm}}}
\begin{table}[t]
\centering
\begin{tabular}{ L C C }
\toprule
	\small\bf Method & \small\bf Gas cost & \small\bf USD cost\\
\midrule
(contract creation) & 358,600 & \$0.46 \\ 
\textsf{initChannel} & 81,053 & \$0.10 \\ 
\textsf{closeChannel} & 114,757 & \$0.15 \\  
\textsf{channelTimeout} & 21,732 & \$0.03 \\
\bottomrule
&&\\
\end{tabular}
\caption{A table showing the gas costs of executing the methods of the payment channel smart contract. The USD costs, which is pegged to the price of Ethereum, are accurate as of April 26, 2018 and assume a gas price of 2 Gwei ($2\cdot 10^{-9}$ Ether). For $closeChannel$, the figures are for calling $closeChannel$ twice, for each respective signature.}\label{table:contractcosts}
\end{table}

\section{Related Work}
\label{sec:related_work}
Result verification has been an active area of research in the past with multiple proposed techniques. The first group of techniques focuses on constructing cryptographic proof of computation \cite{walfish2015verifying, oliaiy2017verifiable, fiore2016hash, wahby2016verifiable}. Such proofs are easy to verify without the need to re-execute the computations. However, the overhead of pre-computation and creation of the proof is orders of magnitude higher than the actual cost of the computation being verified.
The second group of techniques consists of running the same computations on multiple servers \cite{canetti2011practical, distler2016resource, van2014versum, szajda2003hardening}. As long as a given fraction of servers is honest, the result can be guaranteed by a consensus protocol. These techniques require at least one honest server and increase significantly the overhead as they rely on repeating the computation. 

An approach closer to our work involving blockchain technology is \cite{dong2017betrayal}. The authors assume only two execution entities and design smart contracts discouraging them from colluding. Similarly, in \cite{pham2014optimal}, authors determine an optimal penalty fee that should be paid by execution platforms caught cheating. Huang \etal \cite{huang2016bitcoin} also distribute the task to multiple workers and exploit Commitment-based sampling \cite{du2010uncheatable} to verify the correctness of the result. Before starting the computations, the workers have to commit to the task by spending bitcoins that are lost in case of dishonest behaviour. However, both systems (\cite{dong2017betrayal, huang2016bitcoin}) still require to repeat computations and assume a trusted 3rd party to resolve conflicts. 

Multiple projects focus on incentivising fairness and timely delivery of the results using cryptocurrencies \cite{andrychowicz2014secure, bentov2014use, kumaresan2014use, kumaresan2016amortizing, chen2012efficient}. Workers deposit predefined amount of money that is lost if they misbehave. However, all of them focus on fairness exclusively and ignore verifiability of produced results. 
Finally, several projects aim to facilitate blockchain-based micro-payments \cite{lind1612teechan, lundqvist2017thing, poon2015bitcoin}. Those projects are complementary to ours, and could be used to lower the cost and overhead of transactions. 

Several systems provide computations verification, but are limited to a specific class of tasks \cite{shan2018practical}. \cite{carbunar2010fair} supports only tasks that can be easily verified by requesters and has lower security (execution platforms are being fairly paid for their job with a probability < 1) and do not provide privacy of the results. Hu \etal \cite{hu2015secure} focus on both task verification and data privacy, but the system does not include payments and works only for computation of the characteristic polynomial of matrix. 


Several industrial platforms were launched recently aiming to realise a vision of a global decentralised computer. Golem \cite{golem}, focuses mainly on video rendering tasks. Execution platforms are automatically paid if the requester confirms completion of the tasks. However, in case of a conflict, the system relies on a consent that must be trusted by both parties. SONM \cite{sonm} develops a cloud-like services platform based on fog computing as a backend. Execution platforms install SONM OS allowing to share and rent their resources (\ie CPU, GPU, Storage). So far, SONM do not provide any details about their result verification techniques. iExec \cite{iexec}, yet another platform for result verification and automatic payment announced to use Intel SGX in their system, but does not provide any additional details. Finally, Bounty0x \cite{bounty0x} is an open platform allowing people to post tasks with an associated prize. Anyone can then submit a solution and collect the rewards. However, Bounty0x does not provide an automatic verification process and relies on manual conflict resolution.

\section{Conclusion}

We have proposed \sysname, a protocol that \emph{i)} enables requesters to execute tasks on nodes with TEE-enabled CPUs, and \emph{ii)} allows the executing node to receive payments without either party having to trust each other and without the need for a third party to resolve disputes. We employ smart contracts to act as a trustless mediator for the fair exchange of payment and execution result. We also use checkpoint-based micropayments (through payment channels) to reduce the impact of a requester going offline or becoming dishonest after a long execution and wasting the requesting node’s resources without payment.

Our evaluation of Game of Life and OCR show a clear trade-off between efficiency and low-risk. Executing \sysname a few cycles at a time helps with mitigating the risk of requestors dropping out without paying for the execution, but this comes at the cost of both bandwidth and enclave execution time.
\section*{Acknowledgements}

Mustafa Al-Bassam is supported by a scholarship from The Alan Turing Institute, Alberto Sonnino is supported by the EU H2020 DECODE project under grant agreement number 732546, Michał Król is supported by the EC H2020 ICN2020 project under grant agreement number 723014, and Ioannis Psaras is supported by the EPSRC INSP Early Career Fellowship under grant agreement number EP/M003787/1.

Many thanks to Patrick McCorry for discussions about the \sysname design.

\bibliography{biblio} 
\bibliographystyle{ACM-Reference-Format}

\end{document}